\documentclass[]{spie}  

\usepackage{amsmath,amsfonts,amssymb}
\usepackage{graphicx}
\usepackage[colorlinks=true, allcolors=blue]{hyperref}
\usepackage{float}

\title{Bi-objective optimization of organ properties for the simulation of intracavitary brachytherapy applicator placement in cervical cancer}

\usepackage{tikz,xcolor,hyperref}

\definecolor{lime}{HTML}{A6CE39}
\DeclareRobustCommand{\orcidicon}{
	\begin{tikzpicture}
	\draw[lime, fill=lime] (0,0) 
	circle [radius=0.16] 
	node[white] {{\fontfamily{qag}\selectfont \tiny ID}};
	\draw[white, fill=white] (-0.0625,0.095) 
	circle [radius=0.007];
	\end{tikzpicture}
	\hspace{-2mm}
}

\foreach \x in {A, ..., Z}{%
	\expandafter\xdef\csname orcid\x\endcsname{\noexpand\href{https://orcid.org/\csname orcidauthor\x\endcsname}{\noexpand\orcidicon}}
}

\author[a]{Cedric J. Rodriguez\orcidA{}}
\author[a]{Stephanie M. de Boer\orcidD{}}
\author[b]{Peter A. N. Bosman\orcidC{}}
\author[a]{Tanja Alderliesten\orcidB{}}
\affil[a]{Dept. of Radiation Oncology, Leiden University Medical Center (LUMC), P.O. Box 9600, 2300 RC Leiden, The Netherlands}
\affil[b]{Evolutionary Intelligence Group, Centrum Wiskunde \& Informatica (CWI), P.O. Box 94079, 1090 GB Amsterdam, The Netherlands}

\authorinfo{Send correspondence to: \\
Tanja Alderliesten. Email: T.Alderliesten@lumc.nl\\
Cedric Rodriguez. E-mail: C.Rodriguez@lumc.nl}
\begin{document} 
\maketitle

\begin{abstract}
Validation of deformable image registration techniques is extremely important, but hard, especially when complex deformations or content mismatch are involved. These complex deformations and content mismatch, for example, occur after the placement of an applicator for brachytherapy for cervical cancer. Virtual phantoms could enable the creation of validation data sets with ground truth deformations that simulate the large deformations that occur between image acquisitions. However, the quality of the multi-organ Finite Element Method (FEM)-based simulations is dependent on the patient-specific external forces and mechanical properties assigned to the organs. A common approach to calibrate these simulation parameters is through optimization, finding the parameter settings that optimize the match between the outcome of the simulation and reality. When considering inherently simplified organ models, we hypothesize that the optimal deformations of one organ cannot be achieved with a single parameter setting without compromising the optimality of the deformation of the surrounding organs. This means that there will be a trade-off between the optimal deformations of adjacent organs, such as the vagina-uterus and bladder. This work therefore proposes and evaluates a multi-objective optimization approach where the trade-off between organ deformations can be assessed after optimization. We showcase what the extent of the trade-off looks like when bi-objectively optimizing the patient-specific mechanical properties and external forces of the vagina-uterus and bladder for FEM-based simulations. 
\end{abstract}


\begin{figure}[H]
    \centering
    \includegraphics[width=0.9\linewidth]{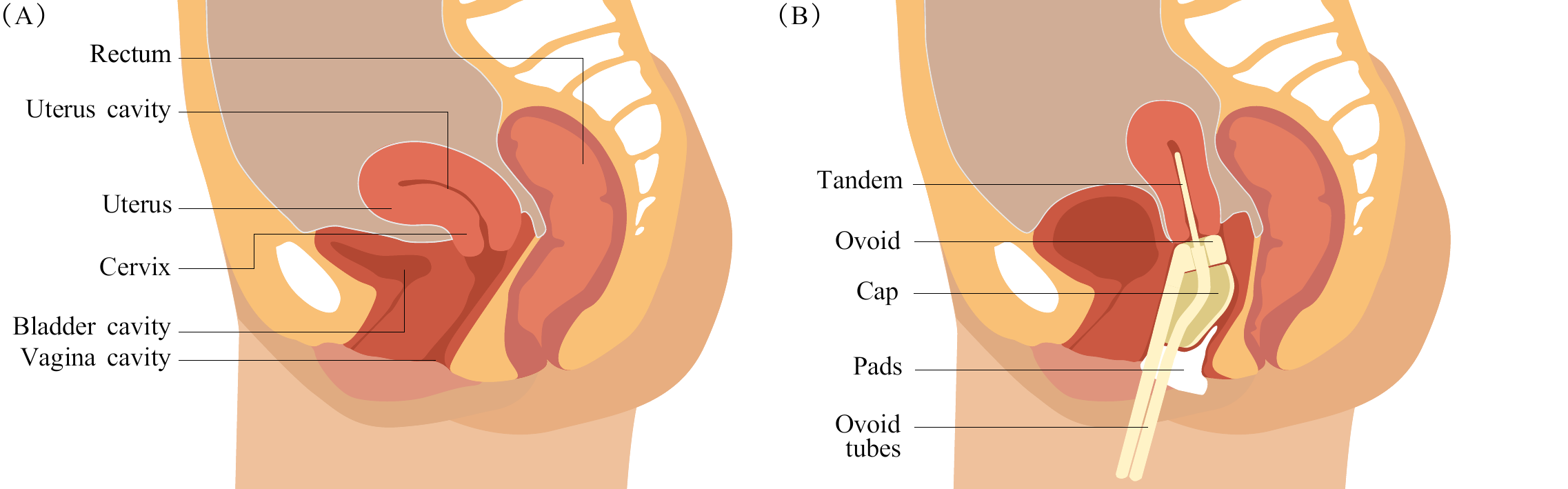}
    \caption{(A) Anatomical location of the vagina, cervix, uterus, and bladder. (B) The placement and main components of an applicator, such as the tube-shaped tandem and the lunar or capsule-shaped ovoids, used for the localized delivery of radiation dose in BT.}
    \label{fig:anatomy}
\end{figure}

\section{INTRODUCTION}
\label{sec:introduction}
Standard treatment for women with locally advanced cervical cancer consists of external beam radiation therapy (RT) with concurrent platinum-based chemotherapy followed by Magnetic Resonance Imaging (MRI)-guided brachytherapy (BT).\cite{cibula2018european} Deformable image registration could play a key role here by finding accurate spatial mappings (tissue correspondence) between the external beam RT treatment planning images and the BT treatment planning images of the abdominal region. The anatomy of the relevant organs during treatment are depicted in Figure \ref{fig:anatomy}. Accurate spatial mappings could potentially be used to transfer the previously delivered dose distributions induced by the external beam RT treatment to the BT treatment planning images, such that previous delivered tumour and organ dosages (such as the bladder and rectum) could be considered during the planning of the BT treatment. Finding an accurate spatial mapping is particularly challenging in the case of cervical cancer due to the large natural variability of the internal organs of the abdominal region over time, such as uterine displacements and the filling of the bladder and rectum.\cite{jadon2014systematic} Additionally, the intracavitary placement of an applicator that is used for delivery of the radiation dose, as depicted in Figure \ref{fig:anatomy}, causes very large deformations as well as a large content mismatch when comparing the images acquired for BT planning to the images acquired for diagnostic purposes and external beam RT planning. \\
\indent Wide use of deformable image registration techniques in clinical practice is still hindered due to a lack of relevant ground truth data sets. Potentially, this could be overcome with physics-based virtual phantoms that ensure physically plausible deformations. Many publications acknowledge the importance of the organ mechanical properties, fixations, and external forces, for the final physics-based deformation due to the patient-to-patient variability. Yet, most of these works use generic mechanical properties based on literature.\cite{bhattarai2018modelling, brock2005accuracy, chanda2016biofidelic, courtecuisse2020three, diallo2022simulation, dias2017pelvic, peng2016assessment, rigaud2019deformable, venugopala2010experiments} A potential solution is to explicitly optimize these properties on a patient-to-patient basis. Some work has attempted to single-objectively optimize the elastic properties of the vagina or muscles of the pelvic floor using multi-organ Finite Element Method (FEM) simulations of the pelvic region, showing promising results. \cite{silva2019characterizing, mayeur2019patient}\\
\indent In this work, we consider finding the simulation parameter settings such that there is an optimal deformation of \textit{both} the vagina-uterus \textit{and} the bladder between diagnostic MRI scans and BT planning MRI scans. This is challenging because these organs are directly adjacent in the abdominal region. Unless we have the perfect simulation setup in which everything that is relevant is modelled correctly, we hypothesize that the optimal deformation of both organs cannot be acquired simultaneously. Hence, there might not be a single Utopian simulation parameter setting that maximizes the deformations of both organs that are optimal within the limits of the employed model. We therefore propose to take a bi-objective approach where one objective is to acquire the optimal deformation of the vagina-uterus and the other objective is to acquire the optimal deformation of the bladder. In contrast to a single-objective approach, a bi-objective optimization approach results in an approximation of, \textit{not one} optimal setting of simulation parameters, but \textit{a set} of simulation parameter settings (set of solutions). This optimal set is also known as the Pareto set. This solution set, by definition, only contains non-dominated solutions. A solution is considered non-dominated when no other solution exists with better objective values in all the objectives simultaneously. The objective values of each solution in the Pareto set form the Pareto front, representing the different optimal trade-off solutions between the best fitting deformation of the vagina-uterus and the deformation of the best fitting bladder. This set provides insight in the range of most plausible organ deformations, given the fidelity of the simulation. A clinical expert can then decide which trade-off is considered most plausible or relevant.\\ 
\indent In this work, we are the first to approach the problem of patient-specific tuning of organ properties and external forces for simulation purposes using a multi-objective optimization approach. Specifically, we apply an Evolutionary Algorithm (EA) to bi-objectively optimize ten pre-selected mechanical properties of the vagina-uterus and bladder to achieve physically plausible organ deformations using FEM-based simulations of the placement of an applicator that is used for delivery of the radiation dose in BT for cervical cancer. This allows us to quantitatively and qualitatively investigate what the extent of the trade-off between the vagina-uterus and the bladder deformations looks like.
\clearpage

\begin{figure}
  \includegraphics[width=0.97\textwidth]{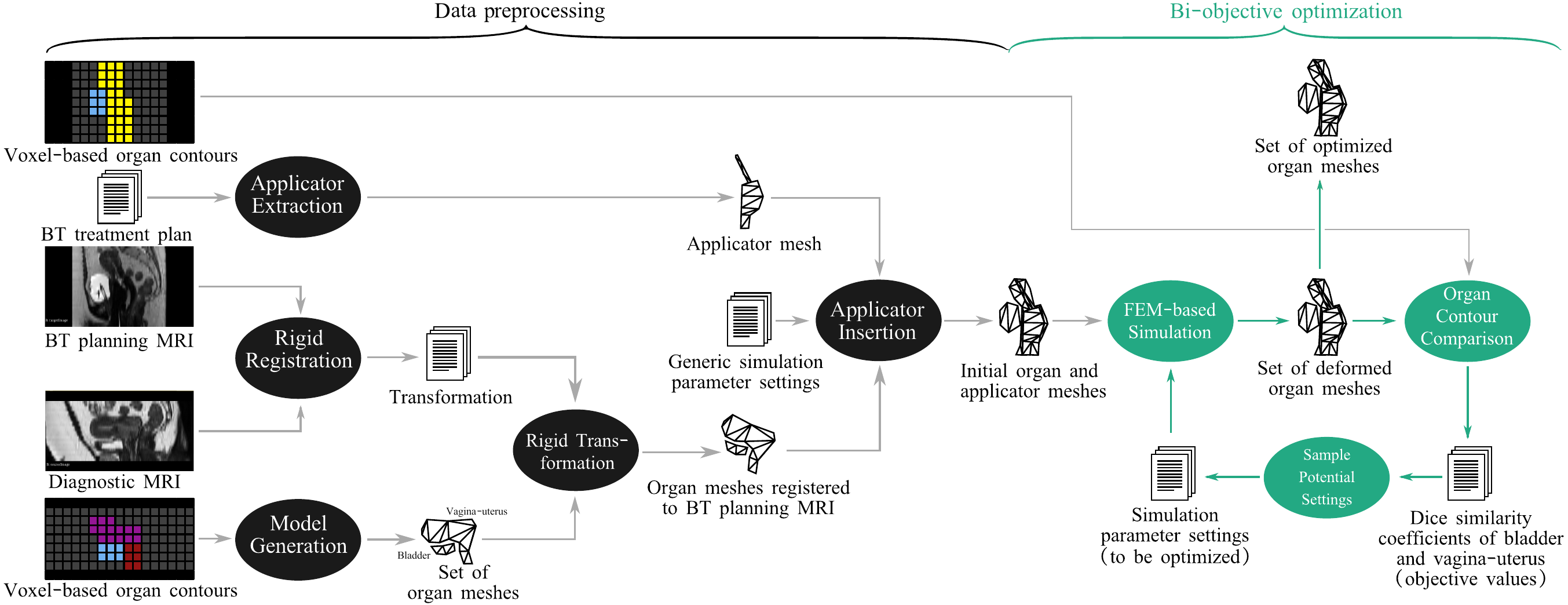}
  \caption{Pipeline for the optimization of the simulation parameters (i.e., mechanical properties and external forces of the vagina-uterus and bladder) to improve the match between simulated and real deformations resulting from applicator placement for brachytherapy (BT) for cervical cancer.}
  \label{fig:teaser}
\end{figure}

\section{OPTIMIZATION PIPELINE}
\label{sec:materials-and-methods}
\subsection{Data}
We retrospectively used T2-weighted MRI scans of two patients treated with BT for cervical cancer. All scans are acquired using the 1.5T Ingenia MRI scanner (Philips Healthcare, Best, the Netherlands). The data of both patients consists of an MRI scan acquired for diagnostic purposes. The data of patient one additionally consists of two BT planning MRI scans acquired after two separate applicator placements. During the first placement, the Venezia gynaecological applicator (Elekta, Veenendaal, the Netherlands) was used with an ovoid size of 26$mm$ without a cap. During the second placement, the Utrecht gynaecological applicator (Elekta, Veenendaal, the Netherlands) was used with an ovoid size of 20$mm$. During both placements, a tandem length of 60$mm$ at 30$^{\circ}$ was used. The data of patient two consists of one BT planning MRI scan acquired after one applicator placement. Patient two was treated using the Venezia gynaecological applicator with an ovoid size of 22$mm$ with a cap. A tandem length of 60$mm$ at 30$^{\circ}$ was used. In all images, the uterus, cervix, vagina, bladder, and abdomino-pelvic cavity were contoured for this study by a radiation therapy technologist and checked by a radiation oncologist.  In this work, the vagina, cervix, and uterus were delineated as one anatomical structure. We therefore denote this structure, that starts from the vagina and ends at the uterus, as the vagina-uterus.

\subsection{Data preprocessing}
\label{subsec:preprocessing}
\begin{figure}[H]
    \centering
    \includegraphics[width=0.95\linewidth]{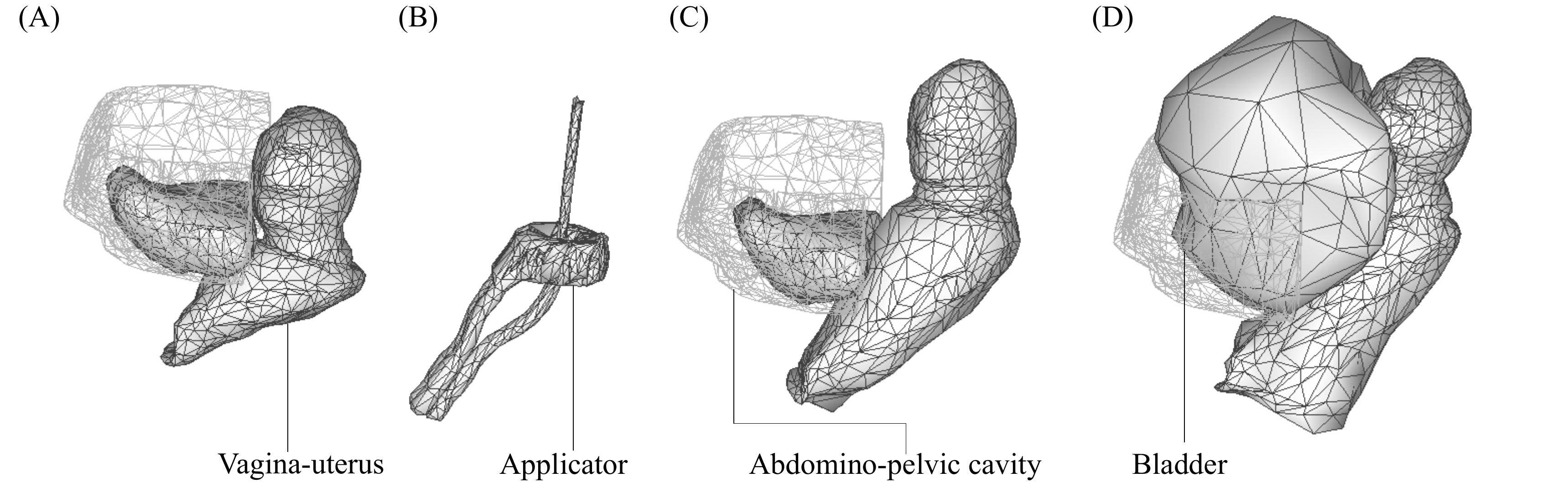}
    \caption{(A) Mesh of the vagina-uterus, bladder, and abdomino-pelvic cavity after the model generation process but before the applicator insertion process of placement 1 of patient 1. (B) Applicator mesh after the applicator extraction process. (C) Mesh shape of the vagina-uterus, bladder, abdomino-pelvic cavity, and applicator after the applicator insertion process but before the bi-objective optimization step. The applicator mesh is contained inside the vagina-uterus mesh. (D) Example organ mesh shape of the vagina-uterus, bladder, abdomino-pelvic cavity, and applicator at the end of the bi-objective optimization step.}
    \label{fig:meshes}
\end{figure}

\indent Figure \ref{fig:teaser} shows the pipeline proposed in this work. The first step in the pipeline consists of a data preprocessing step which converts the organ contour data into mesh representations. A \textit{model generation} process converts the contours of the organs from a 3D binary label map based on a diagnostic MRI scan into a tetrahedron mesh. The resulting meshes are then transformed to the desired coordinate system of the target MRI based on translations and rotations of a \textit{rigid registration} and \textit{rigid transformation} process. This process aligns the bony anatomy of the patient and is used as the initial alignment of the diagnostic MRI scan and the BT planning MRI scan for the bi-objective optimization. The applicator geometry is extracted from the treatment plan and is then converted into a surface mesh. In the \textit{applicator insertion} process, we  simulate the initial deformation of the vagina-uterus by inserting the applicator mesh inside the uterine cavity while assuming generic simulation parameters as described in Section \ref{sec:simparams}. The lower bound of the literature reported parameters ranges are used, which corresponds to low forces and easily deformable tissue. To ensure stable bases of the organs, the 20$mm$ and 10$mm$ most inferior nodes of the vagina-uterus and bladder meshes, respectively, are constrained (i.e., fixed) in the superior-inferior direction. The abdomino-pelvic cavity and the applicator are considered to be rigid and fixed objects. All the details of the data preprocessing step can be found in Appendix \ref{appendix:preprocessing}. Further, example mesh representations at different stages are illustrated in Figure \ref{fig:meshes}.

\subsection{Bi-objective optimization}
\indent We have chosen to use an EA as the optimization algorithm because they are naturally suited for multi-objective optimization.\cite{deb2001multi} We specifically consider a state-of-the-art multi-objective estimation-of-distribution algorithm called MAMaLGaM. We provide a brief outline of its key components, but refer the interested reader for all details to the literature.\cite{bosman2010anticipated} Similar to many EAs, MAMaLGaM starts with a randomly generated set (population) of simulation parameter settings (solutions). For each solution in the population, a complete simulation is executed to determine a solution's quality (called an evaluation). The resulting quality score for the bladder and the vagina-uterus is then assigned to each solution (simulation parameter setting). The algorithm iteratively (through generations) selects the best solutions and generates new solutions for the population of the next generation. Typically, the population size, maximum number of generations, and maximum number of evaluations of the solutions are user-defined and problem-specific. Within each generation, MAMaLGaM uses two methods in sequence to divide the solutions into a predefined $K$ number of clusters. For the first $D$ clusters, where $D$ is the number of problem dimensions, the best $N_D$ solutions within each problem dimension are grouped together. Before applying the second clustering method, all remaining solutions are ranked based on domination, where the first rank contains all non-dominated solutions and the subsequent ranks are all the non-dominated solutions after removing the lower/previous ranks. Subsequently, a fraction $\tau$ of the population (the best-ranked solutions) is clustered into the remaining $K - D$ clusters using balanced k-leader-means clustering in the objective space.\cite{rodrigues2014novel} A Gaussian distribution is estimated for each cluster and new offspring solutions are sampled from these distributions to form the population of the next generation. These offspring solutions can then be evaluated by running a simulation for each new solution (simulation parameter setting). In our application, MAMaLGaM is terminated when the predefined maximum number of simulations has been reached. 

\subsubsection{Simulation parameter settings}
\label{sec:simparams}
To approximate the influence of the surrounding tissue on the vagina-uterus, a constant external force is added to all the nodes of the vagina-uterus mesh, which introduces three to be optimized simulation parameters (one value per orthogonal direction) initialized between -1 and 1$kN$ which corresponds to a gravitational force of 100$kg$. The resulting internal forces inside the tissue are governed by, in our case, linear elastic equations. This relation is governed by the elastic modulus (where a higher modulus means stiffer material) of the vagina-uterus and is a simulation parameter initialized between 0.001 and 100$N/mm^2$.\cite{baah2016mechanical} To ensure the fixation of the applicator in clinical practice, the vagina is filled with pads, which results in an expansion of the vagina. To model this effect while limiting the number of parameters in our simulation, four sphere shaped objects are placed inside the cavity of the vagina. The diameters of these spheres model the amount of padding inserted in the vagina and are initialized between 0 and 40$mm$ corresponding to the size of the applicator. This results in a total of eight simulation parameters to be optimized for the vagina-uterus mesh. The bladder is modelled as a hollow organ with an inner and outer surface. The internal forces are modelled using a constant elasticity between 0.05 and 11.5$N/mm^2$.\cite{martins2011uniaxial} Further, the target bladder volume, which is filled using a Foley balloon catheter before acquiring the BT planning MRI scans, is considered to be a simulation parameter to be optimized between 0 and 600$cm^3$.\cite{pos2003influence}

\subsubsection{Simulation evaluations}
Simulations are executed using the FEM open-source framework SOFA (v20.12.02). The simulation parameters are empirically defined. The simulation is divided in 2 $\cdot$ 10\textsuperscript{3} simulation steps with an internal step size of 2.5 $\cdot$ 10\textsuperscript{-3} $s$ (i.e., simulating 5$s$ in simulation time) such that the simulation is likely to be numerically stable while limiting the number of simulation steps. The simulation consists of four meshes: the vagina-uterus mesh, the bladder mesh, the applicator mesh, and the abdomino-pelvic  cavity mesh, see Figure \ref{fig:meshes}. The latter is used to constrain the amount of anterior and lateral movements of the bladder. Both the abdomino-pelvic cavity and applicator meshes are rigid triangular surface meshes, since their elasticity is much higher than the elasticity of the vagina-uterus and bladder. After each simulation, the deformation quality (i.e., the objective values during the bi-objective optimization step) has to be quantified for each organ separately. The quality of the deformed organs is assessed by comparing the associated deformed organ contours to the BT treatment planning contours using the Dice similarity metric.\cite{dice1945measures} Some solutions correspond to simulation failure, e.g., due to implausible parameter combinations or fast dynamical behaviour, which result in unstable simulations. In such cases, the worst possible Dice similarity score of zero is returned for both organs. This is also done when the bladder and vagina-uterus tissue are intersecting.

\section{EXPERIMENTS}
To test that the optimal deformation of both organs cannot be acquired simultaneously, the complete pipeline is executed on three BT applicator placements. Literature-based settings for MAMaLGaM have been used.\cite{bosman2009empirical}  In this case study, the number of simulations that the EA could execute is limited (3000 evaluations) due to the long simulation time involved with each evaluation (20 minutes). Since MAMaLGaM is a stochastic optimization algorithm, the optimization is repeated 20 times per applicator placement to make sure the attained results are robust. For each applicator placement separately, the approximated Pareto sets and fronts of all runs are aggregated. The quality of the optimization is evaluated using an independent quality metric (root mean square). The root mean square metric indicates the average shortest distance between the two organ surfaces and is calculated using the open-source 3D mesh processing software MeshLab (v2020.12). To calculate the average distance, $2 \cdot 10^{5}$ points are randomly sampled on the optimized organ mesh. Subsequently, the shortest squared distance is determined to the target organ surface. By averaging over all these points and taking the square root, the final metric score is determined. Finally, the results are also assessed qualitatively by comparing the optimized organ contours to the target organ contours. Not all deformations can be visualized simultaneously. We therefore decided to use a subset of 5 solutions that are distributed along the approximated Pareto front. The subset of the solutions per placement is determined using a greedy hypervolume subset selection method.\cite{guerreiro2016greedy} As the name suggests, this method uses the (to be maximized) hypervolume which is a performance indicator in the case of multi-objective optimization algorithms. Hypervolume defines the space enclosed between the Pareto front and a user-defined reference point. The method sequentially selects the solution from the approximated Pareto set that has the largest contribution to the hypervolume. This typically will lead to the selection of the solutions that mostly characterize the approximated Pareto front.

\section{RESULTS AND DISCUSSION}
\label{sec:results-and-discussion}
\begin{figure}[H]
    \centering
    \includegraphics[width=\linewidth]{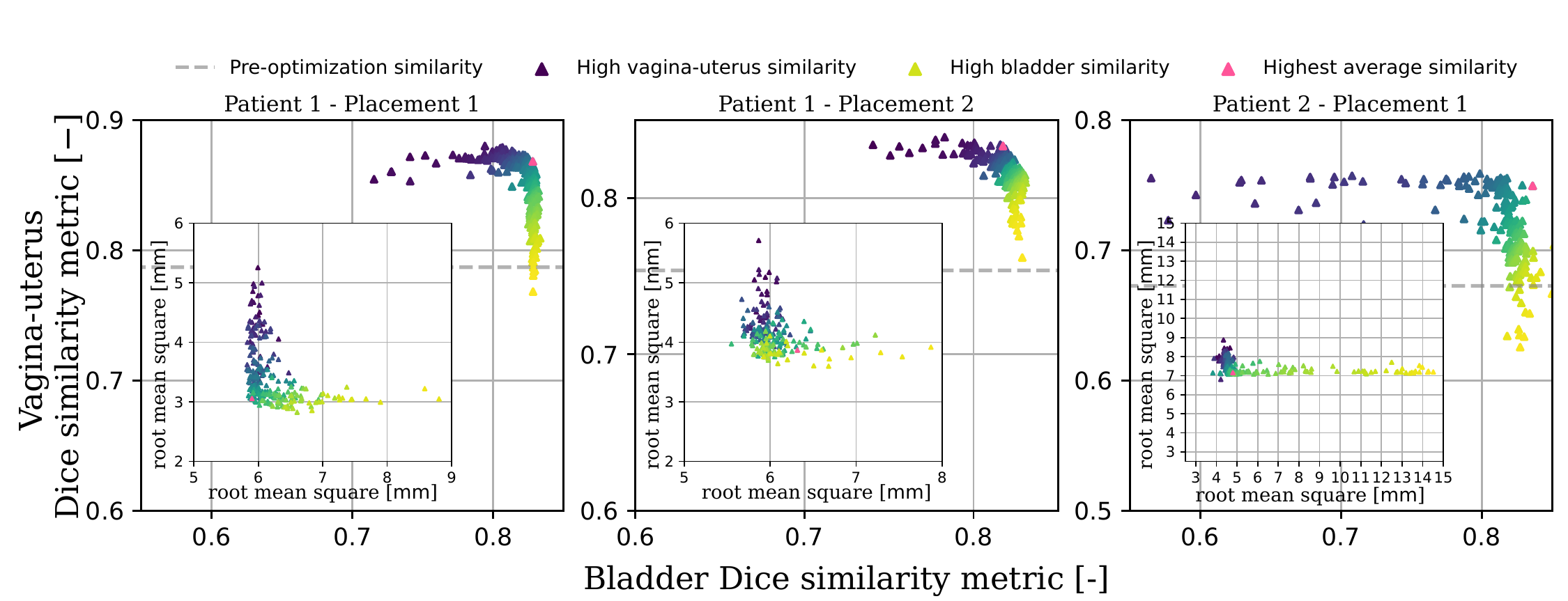}
    \caption{Aggregation of the approximation fronts of twenty independent runs using the Dice similarity score of the vagina-uterus and bladder as the two objectives during optimization of the simulation parameter settings. The solution coloured in pink is the solution with the highest average Dice similarity score of the vagina-uterus and bladder. Pre-optimization similarity refers to the Dice similarity score after the data preprocessing step but before the bi-objective optimization step. For all three placements, the pre-optimization similarity scores of the bladder  of 0.27, 0.33, and 0.22 are outside the plotting range. The subplots depict the (independent) similarity scores using the root mean square metric.}
    \label{fig:approxSet}
\end{figure}

Figure \ref{fig:approxSet} depicts the aggregation of all approximated Pareto fronts resulting from the twenty individual optimization runs. Higher Dice similarity scores are found than what is observed for the initial organ shapes after the data preprocessing step, but before the bi-objective optimization step. This shows that there definitely is a benefit to optimizing the simulation parameter settings. We find a trade-off front instead of one single Utopian solution for each applicator placement. The trade-off persists, even when evaluating the deformation quality with another metric (i.e., root mean square of the distances between the organ surfaces). \\
\indent In contrast to multi-organ optimization, single-objective approaches typically require the selection of a similarity metric that aggregates different optimization objectives (e.g., the average Dice similarity score of all organs). After optimization, this will result in one simulation parameter setting which maximizes the average Dice similarity score. This solution is coloured in pink in Figure \ref{fig:approxSet}. Since the metric is defined a priori, and since there clearly is a trade-off between the quality of deformations, this leaves the end user uninformed about all the (potentially equally relevant) organ deformations, and it may lead to wrong conclusions about the capacity of the simulation. In some clinical cases, it could for example be appropriate to prioritize the deformation quality of the vagina-uterus. Taking a bi-objective approach makes it possible to a posteriori choose another solution along the approximation front, i.e., one that has a higher similarity score for the vagina-uterus at the cost of a lower similarity score for the bladder.\\

\begin{figure}[H]
    \centering
    \includegraphics[width=\linewidth]{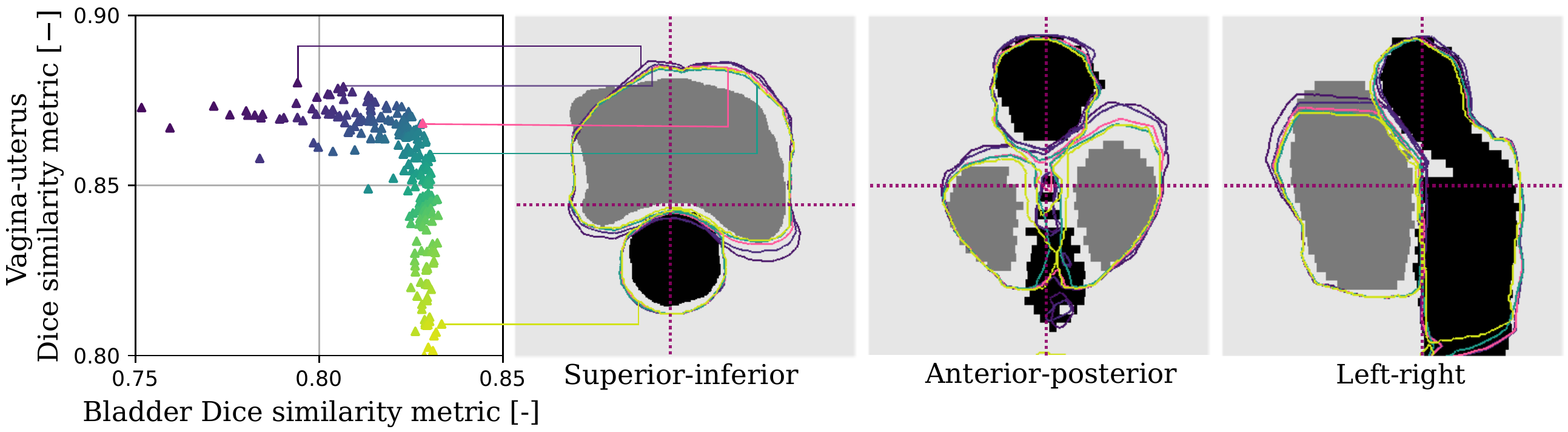}
    \caption{Orthogonal views of patient 1 (placement 1) with the target bladder (grey), target vagina-uterus (black), and the solution contours  of five solutions selected from the approximation front. The same colour coding of solutions is used as in Figure \ref{fig:approxSet}.}
    \label{fig:deformations11}
\end{figure}

\begin{figure}[H]
    \centering
    \includegraphics[width=\linewidth]{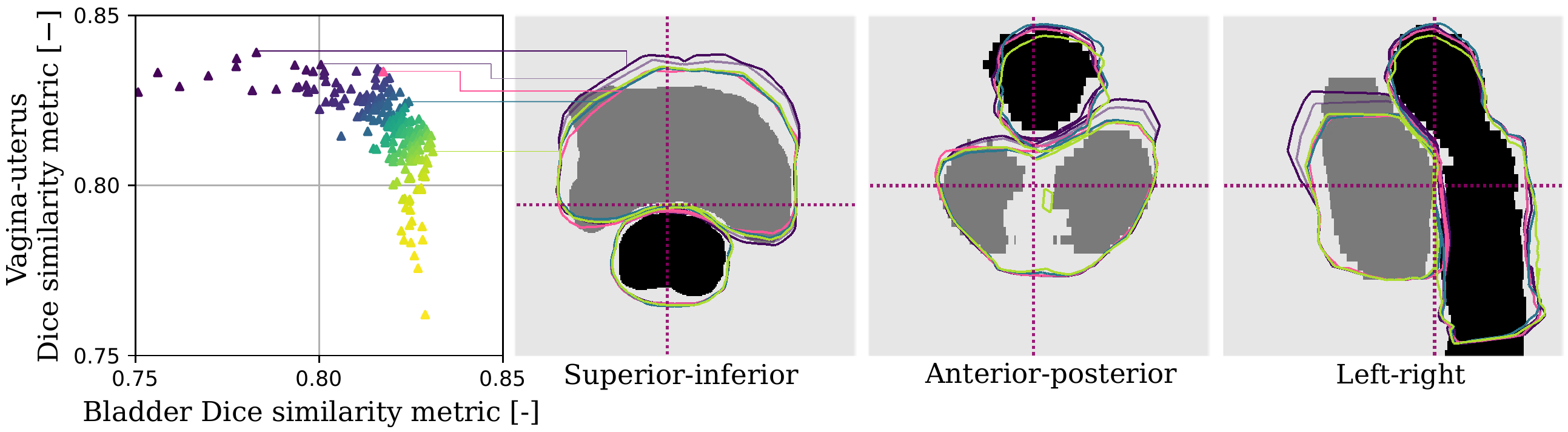}
    \caption{Orthogonal views of patient 1 (placement 2) with the target bladder (grey), target vagina-uterus (black), and the solution contours  of five solutions selected from the approximation front. The same colour coding of solutions is used as in Figure \ref{fig:approxSet}.}
    \label{fig:deformations13}
\end{figure}

\begin{figure}[H]
    \centering
    \includegraphics[width=\linewidth]{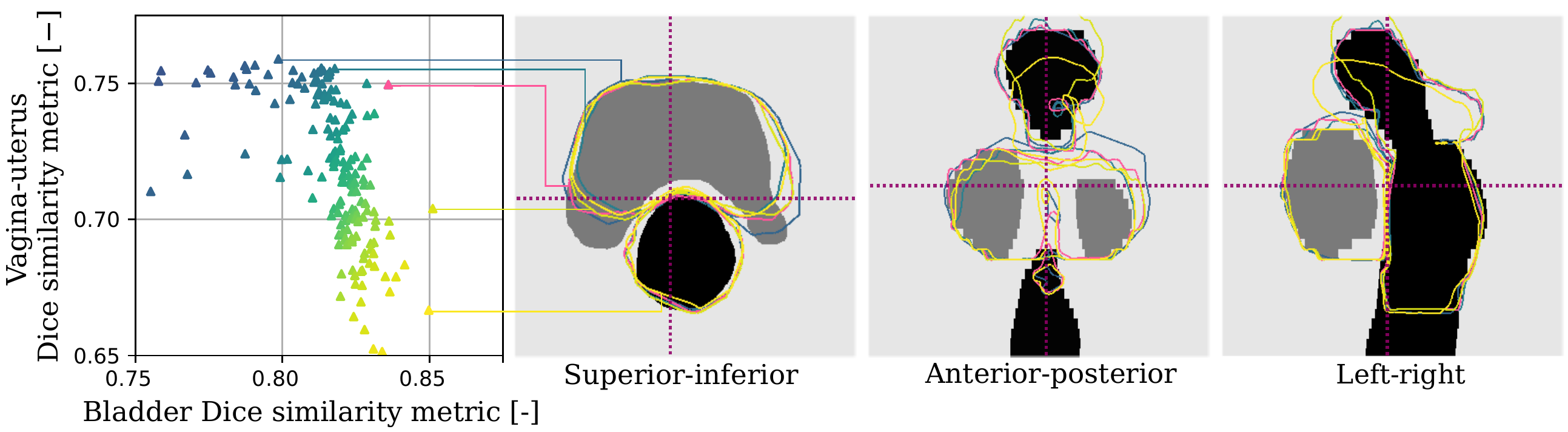}
    \caption{Orthogonal views of patient 2 (placement 1) with the target bladder (grey), target vagina-uterus (black), and the solution contours  of five solutions selected from the approximation front. The same colour coding of solutions is used as in Figure \ref{fig:approxSet}.}
    \label{fig:deformations21}
\end{figure}

\indent To visualize the deformed organs pertaining to the solutions along the aggregated approximation front, a set of five solutions is selected, see Figure \ref{fig:deformations11}, \ref{fig:deformations13}, and \ref{fig:deformations21}. In the left-right view of Figure \ref{fig:deformations11} and \ref{fig:deformations13}, it shows that reducing the bladder volume and simultaneously applying a force in the inferior direction on the vagina-uterus increases similarity of the bladder and dissimilarity of the vagina-uterus at the same time. Likewise, increasing the bladder volume leads to improved vagina-uterus deformations but results in an abnormal increase in the bladder in the anterior and left-right direction. Taking a bi-objective approach makes these trade-offs explicit, enabling a clinical expert to select which deformation is considered most plausible or relevant. It is not surprising that the trade-off in the organ deformations illustrated in Figure \ref{fig:deformations11} and \ref{fig:deformations13} are similar, since these concern optimization results obtained for the same patient and are based on the same diagnostic scan derived organ meshes (and their limitations). Although the trade-offs look similar, they are not exactly the same. The difference in the highest Dice similarity score obtained for the vagina-uterus between placement 1 and 2 is 0.035. We expect that this has to do with the unrealistic deformation of the optimized vagina-uterus mesh near the edge of the field-of-view of the scans. Since the vagina-uterus is undefined outside the field-of-view of the diagnostic MRI scans, this leads to abnormal deformation around the inferior slices of the MRI scans. These slices are clinically less relevant for the registration in our application, since the gradients of the radiation dose distributions are relatively small compared to those in the cervical region.

Figure \ref{fig:deformations21} shows the trade-off between the deformation of the vagina-uterus and the bladder of placement 1 of patient 2. In the left-right view, the simulation parameter setting with the highest achievable bladder Dice similarity score also results in an inferior deformation of the vagina-uterus. Although, during this application, this inferior deformation is more extreme compared to patient 1. When looking at the anterior-posterior view, it appears that the vagina-uterus is used to enforce an asymmetrical deformation of the bladder. Furthermore, the application of patient 2 has an overall lower vagina-uterus Dice similarity score than the applications of patient 1. This is partly due to the larger mismatch in the field-of-view between the diagnostic MRI scan and the BT planning MRI scan. The second reason for lower maximum achievable Dice similarity scores is due to the large mismatch in tumour size between the MRI scans of patient 2. The organ models are based on diagnostic MRI scans which have been acquired prior to external beam RT with concurrent platinum-based chemotherapy. During the treatment, large tumour regression can occur. Only applying a global force onto the vagina-uterus cannot lead to this local shrinkage of the tumour inside the vagina-uterus. 

Although these results are promising, several limitations of the work have to be discussed. In this problem formulation, only two organs are considered and no inter organ tissue is considered. Including additional abdominal organs (i.e., bowel or rectum) or other structures (i.e., pelvis) should improve the clinical relevance of the simulations. Moreover, future organ models should be able to consider the tumour regression between scans or additional image acquisition directly prior to the applicator placement could be implemented. By subsequently increasing the number of patients, in-depth analyses would be possible of the simulation parameter settings within (and between) different patients. Furthermore, the field-of-view of the simulated organs should match the field-of-view of the BT planning MRI and abnormal deformation near the edge of the field-of-view of the simulation should be addressed. Even though key simulation parameters settings have been selected, it is unclear how sensitive the estimated external forces and mechanical parameter values are to the problem formulation (i.e., organ fixations).\\
\indent The bi-objective optimization step currently takes seven days due to the computationally expensive FEM-based simulations. To evaluate one potential setting of the simulation parameters, one full simulation is required. It could be interesting to experiment with expensive optimization techniques where the expensive-to-evaluate simulations are interchanged with a quick-to-evaluate model of the simulation such that less expensive-to-evaluate simulations are required.\cite{wang2006review} This should result in high optimization quality in shorter optimization time.\\
\indent Finally, improved constraint handling could be included in the optimization. The violations of some constraints, such as overlapping organs meshes, could for instance not be penalized with the worst possible fitness values. Instead, a degree of the violations could be considered. Furthermore, the balance between model resolution, simulation time, and simulation numerical stability should be considered. On the one hand, higher model resolutions lead to longer simulation time if the same numerically stability is desired. On the other hand, reducing the model resolution leads to shorter simulation time and improved numerically stability, but obviously leads to reduced simulation fidelity. 

\section{CONCLUSION}
\label{sec:conclusion}
This work shows that there is a trade-off between the deformations of inherently simplified adjacent organ models in the abdominal region in FEM-based simulations of the placement of the applicator for BT for cervical cancer. Understanding and acknowledging this is important, since any a priori single-objective optimization approach may lead to wrong conclusions about the parameters, found for the underlying biomechanical models. By taking a multi-objective approach using an EA, we could approximate the inherent trade-off, enabling the possibility to a posteriori evaluate which of the deformations is considered most plausible by a clinical expert. Our results make explicit the extent of the inherent trade-off between the similarity metrics and their associated deformations of the vagina-uterus and bladder due to inherent simulation model inaccuracies, while at the same time opening the possibility to a posteriori selection of the most plausible organ deformations, thereby increasing the probability of success in practice. 

\section*{ACKOWLEDGEMENTS}
The authors thank E.C. Harderwijk (Dept. of Radiation Oncology, Leiden University Medical Center, Leiden, The Netherlands) for his contributions to this study regarding the data preparation. We further thank Y. Niatsetski and R. Zinkstok (Elekta, Veenendaal, the Netherlands) for the development of a special module to export the brachytherapy applicator geometry. The research is part of the research programme Open Technology Programme with project number 15586, which is financed by the Dutch Research Council (NWO), Elekta (Elekta Solutions AB, Stockholm, Sweden), and Xomnia (Xomnia B.V., Amsterdam, The Netherlands). Further, the work is co-funded by the public-private partnership allowance for top consortia for knowledge and innovation (TKIs) from the Ministry of Economic Affairs.

\bibliography{report} 
\bibliographystyle{spiebib} 

\clearpage
\appendix 
\section{Details of data preprocessing}
\label{appendix:preprocessing}

In this appendix, the details of the model generation, rigid registration, rigid transformation, applicator extraction, and applicator insertion are presented.
\subsection*{Model generation}
The model generation process uses the clinical expert's manually defined outer contours of the vagina, uterus, bladder, and abdominal cavity to create the tetrahedron meshes. Even though the vagina, uterus, and bladder consist of an internal cavity, only the outer surface of the organs could be contoured by a radiation therapy technologist in diagnostic MRI scans (Table \ref{tab:imageres}) due to the poor contrast between the organ cavity and the organ tissue.

\begin{table}[H]
    \caption{Details of the retrospective T2-weighted MRI scans where L-R denotes the Left-Right direction, A-P denotes the Anterior-Posterior direction, and S-I denotes the Superior-Inferior direction.}
    \centering
    \begin{tabular}{|c|c|c c c|c c c|} \hline
        Patient ID & Purpose scans & \multicolumn{3}{c|}{Number of voxels} & \multicolumn{3}{c|}{Voxel size ($mm$)}\\ \hline
          &            & L-R & A-P & S-I & L-R & A-P & S-I \\ \hline
        1 & Diagnostic & 288 & 288 & 24 & 0.6250 & 0.6250 & 4.0000\\
        1 & Placement 1 & 432 & 432 & 40 & 0.5324 & 0.5324 & 4.0000\\
        1 & Placement 2 & 432 & 432 & 40 & 0.5324 & 0.5324 & 4.0000\\ \hline
        2 & Diagnostic & 768 & 768 & 24 & 0.3125 & 0.3125 & 4.0000\\
        2 & Placement 1 & 432 & 432 & 35 & 0.4167 & 0.4167 & 4.4000\\ \hline
        
    \end{tabular}
    \label{tab:imageres}
\end{table}

To be able to insert the applicator inside the vagina and uterus, requires us to make basic assumptions about the internal cavity of these organs. The cavity of the vagina and bladder are determined by specifying a constant wall thickness of 1$mm$ and 2$mm$, respectively.\cite{oelke2006ultrasound, panayi2010ultrasound} The vagina wall thickness is chosen smaller than what is reported in literature (2.7$mm$) such that the opposing inner wall surfaces of the vagina do not overlap. The uterus cavity is a canal which follows the centreline of the uterus starting from the external os towards the fundus, which is created using the plug-in Vascular Modelling Toolkit (v1.4) of the open-source software platform 3D Slicer (v4.11). The length of the canal is chosen to be equal to the applicator tandem length and radius, which is 60$mm$ and 1.5$mm$ for the considered patients. The vagina and uterus are then merged into one vagina-uterus contour. The contours are converted into triangular meshes using 3D Slicer (v4.11).\\
\indent To reduce the simulation time, the surface mesh resolutions of the vagina-uterus, bladder, and abdominal cavity mesh are reduced (see Table \ref{tab:modelSpecs}) using the well-established quadratic edge collapse decimation method in the open-source 3D mesh processing software MeshLab (v2020.12). The lowest resolution is empirically chosen that still describes the internal cavity of the organs. To ensure proper tetrahedralisation of the surface mesh, all manifold or self-intersecting triangular surfaces are removed. Thereafter, a watertight mesh is created by closing the obtained mesh holes. Finally, the open-source mesh generator GMSH (v4.8.1) is used to generate a 3D tetrahedron mesh of the hollow vagina-uterus and bladder surface meshes using the widely used Delaunay method.\cite{si2006quality}\\
\indent The model generation process thus outputs two hollow tetrahedron meshes of the vagina-uterus and bladder, and one surface mesh of the abdominal cavity.

\begin{table}[H]
\caption{Summary of the organ and applicator mesh resolutions (i.e., number of vertices, triangles, and tetrahedra) resulting from the diagnostic MRI scans and BT treatment plan, respectively, of each patient generated in the model generation process.}\label{tab1}
\centering
\scalebox{0.85}{
\begin{tabular}{| c | c c | c c | c c | c c | c |} \hline
   & \multicolumn{2}{c|}{Vagina} & \multicolumn{2}{c|}{Bladder} & \multicolumn{2}{c|}{Abdominal} & \multicolumn{2}{c|}{Applicator} & \multicolumn{1}{c|}{Applicator} \\
    & \multicolumn{2}{c|}{uterus} & \multicolumn{2}{c|}{} & \multicolumn{2}{c|}{cavity} & \multicolumn{2}{c|}{Placement 1} & \multicolumn{1}{c|}{Placement 2} \\ \hline
 Patient & 1 &  2 &  1 & 2 & 1 & 2 & 1 &  2 &  1 \\ \hline
 Vertices & 1345 & 1385 & 502 & 502 & 490 & 499 & 1083 & 923 & 1238\\
 Triangles & 2488 & 2495 & 1000 & 1000 & 2338 & 976 & 2500 & 2500 & 2499\\
 Tetrah. & 4726 & 4683 & 1564 & 1597 & - & - & - & - & -\\ \hline
\end{tabular}}
\label{tab:modelSpecs}
\end{table}

\subsection*{Rigid registration \& model transformation}
\indent The intensity-based rigid registration process is based on the open-source Simple ITK library (v2.1.1) which finds the initial 3D translations and rotations of the diagnostic MRI scans with respect to the BT planning MRI scans. Mutual information with 50 histogram bins combined with a linear interpolator is used as evaluation metric to quantify the similarity between the voxels intensities. The translations and rotations can then be optimized using the Powell optimizer using Brent line search \cite{powell1964efficient} with the default settings of 100 iterations, 100 maximum line iterations, a step length of 1, step tolerance of 1 $\cdot$ 10\textsuperscript{-6}, and value tolerance of 1 $\cdot$ 10\textsuperscript{-6}. This gradient-free optimizer acquired a visually satisfactory alignment of the pelvic bones on this set of patients, but it is possible that other gradient-based Simple ITK methods would work just as satisfactory.\\
\indent Using the acquired 3D rotations and translations, the mesh nodes of the vagina-uterus, bladder, and abdominal cavity can now be transformed from the diagnostic MRI coordinate system to the target BT planning MRI coordinate system.

\subsection*{Applicator extraction}
\indent A large factor influencing the deformation of the vagina-uterus is the applicator position, orientation, and geometry. During daily clinical practice, the applicator is reconstructed in the MRI acquired for BT treatment planning purposes. Using an Oncentra Brachy (v4.6.0, Elekta, Veenendaal, the Netherlands) custom plugin developed by Elekta, the applicator geometry can be extracted in the desired position and orientation as a voxel-based contour. Using 3D Slicer (v4.11), the applicator contour is then converted into a surface mesh. Finally, the resolution of the surface mesh is reduced using the quadratic edge collapse decimation method in MeshLab (v2020.12) to increase the simulation speed. 
\subsection*{Applicator insertion}
To combine the vagina-uterus and applicator mesh, the vagina-uterus is pre-deformed such that the applicator fits inside the organ. To accomplish this, an FEM-based simulation is performed with generic mechanical parameters namely an elastic modulus of the vagina-uterus of $0.001N/mm^2$, elastic modulus of the bladder of $0.05N/mm^3$, external forces of $0kN$ in all directions, sphere diameters of $0mm$, and bladder volume of $0cm^3$. The FEM open-source framework SOFA (v20.12.02) is utilized, where the default simulation pipeline is used for 100 simulation seconds with a step size of 2.5 $\cdot$  10\textsuperscript{-3} $s$. These simulation resolutions ensured stable simulations capturing the (occasionally) fast dynamic oscillations of the organ tissue while reducing the simulation time. For the collision detection between organ surfaces, a brute force detection approach with default contact settings is used. The minimal proximity intersection is used where the collision between elements is detected when they are at an alarm distance of 8$mm$ and make contact at 0.5$mm$ distance.\\
\indent The organ deformations are solved using an Euler explicit solver and a sparse linear solver based on the LDL decomposition.\cite{martin1965symmetric} For the collision models, a triangle, line, and point collision model is used with a contact elasticity equal to the organ elasticity.\\
\indent The applicator cannot directly be placed inside the vagina-uterus mesh due to the initial rest-state of the organ. Placement of the applicator through the vagina entry requires detailed a priori knowledge of all actions (i.e., translations and rotations) applied to the applicator and is not the focus of this work. Rather, we consider finding the end state deformation of the vagina and uterus after the placement of the applicator. Therefore, we first pre-deform the vagina-uterus cavity in the shape of the applicator. To accomplish this, sphere-shaped spacers are generated along the internal cavity path. These spacers are initialized with a diameter of zero and are enlarged linearly in simulation time until the spheres cover most of the applicator target geometry. Simultaneously, the spacers are displaced from the initial vagina-uterus cavity centreline to the target position of the applicator. The spacers form a line from the tip to the base of the tandem and from the base of the tandem to the start of the stem of the applicator, and are evenly distributed along these two lines. After the spacers reach their target size/position, and as a result the vagina-uterus cavity is enlarged, the spacers are removed and the actual applicator mesh is placed inside the vagina-uterus mesh.

\end{document}